\documentclass[aps,prc,reprint,twocolumn,superscriptaddress,nofootinbib]{revtex4-2}

\usepackage{hyperref}
\usepackage{mathtools}
\usepackage{amsmath}
\usepackage{amssymb}
\usepackage{amsbsy}
\usepackage{bm}
\usepackage{textcomp}
\usepackage[dvips]{graphicx,color}
\usepackage{chngcntr}
\usepackage{tikz}
\usetikzlibrary{shapes.geometric, arrows}
\usepackage{fancyhdr}
\usepackage{float}
\usepackage{dsfont}
\usepackage{ulem}

\newcommand{\SU}[1]{\ensuremath{\mathrm{SU}( #1 )}}

\newcommand{\SO}[1]{\ensuremath{\mathrm{SO}( #1 )}}

\newcommand{\SpR}[1]{\ensuremath{\mathrm{Sp}( #1,\mathbb{R} )}}

\newcommand{\ket}[1]{\ensuremath{\left| #1 \right\rangle}}
\newcommand{\braket}[2]{\ensuremath{\langle #1 | #2 \rangle}}

\newcommand{\hw}{\ensuremath{\hbar\Omega}}

\begin{document}

%\textit{Ab initio} informed $^{20}$Ne(p, p$\alpha$)$^{16}$O reaction elucidates alpha clustering physics

\title{
\textit{Ab initio} informed $^{20}$Ne(p, p$\alpha$)$^{16}$O reaction \\ elucidates the emergence of alpha clustering from chiral potentials 
}
\author{G. H. Sargsyan}
\affiliation{Facility for Rare Isotope Beams, Michigan State University, East Lansing, Michigan 48824, USA}
\author{Kazuki Yoshida}
%\email{Present address: Research Center for Nuclear Physics (RCNP), Osaka University, Ibaraki, Osaka, 567-0047, Japan; Interdisciplinary Theoretical and Mathematical Sciences Program (iTHEMS), RIKEN, Wako 351-0198, Japan}
\affiliation{Advanced Science Research Center, Japan Atomic Energy Agency, Tokai, Ibaraki 319-1195, Japan}
\author{Kazuyuki Ogata}
\affiliation{Department of Physics, Kyushu University, Fukuoka 819-0395, Japan}
\affiliation{Research Center for Nuclear Physics (RCNP), Osaka University, Ibaraki, Osaka, 567-0047, Japan}
\author{K. D. Launey}
\affiliation{Department of Physics and Astronomy, Louisiana State University, Baton Rouge, LA 70803, USA}
\author{J. E. Escher}
\affiliation{Lawrence Livermore National  Laboratory, Livermore, California 94550, USA}
\author{D. Langr}
 \affiliation{Department of Computer Systems, Faculty of Information Technology, Czech Technical University in Prague, Prague 16000, Czech Republic}
\author{T. Dytrych}
 \affiliation{Nuclear Physics Institute of the Czech Academy of Sciences, 250 68 \v{R}e\v{z}, Czech Republic}%
 \affiliation{Department of Physics and Astronomy, Louisiana State University, Baton Rouge, LA 70803, USA}

\begin{abstract}
%Targeted journal: Phys Lett B (IF 4.8)
We report on the first \textit{ab initio} informed $\alpha$ knock-out reaction in the intermediate-mass region, with the aim to probe the underlying chiral potential and its impact on the emergence of alpha clustering in this mass region. The theoretical predictions of the $\alpha+^{16}$O clustering in the $^{20}$Ne ground state, based on the \textit{ab initio} symmetry-adapted no-core shell model with continuum, yield a triple differential cross section for $^{20}$Ne(p, p$\alpha$)$^{16}$O  that is in a remarkable agreement with the data. This allows us to examine predictions of surface and in-medium $\alpha$-cluster features that emerge from the underlying realistic nucleon-nucleon interaction with no parameters fitted to nuclear data beyond the two-body system,
%from a chiral potential 
and to compare these to the successful antisymmetrized molecular dynamics approach.

\end{abstract}

\maketitle

\section{Introduction }
%\red{[TBD]} \\

Elucidating $\alpha$ clustering, along with the inter- and intra-cluster structure, is of utmost significance to further understanding dominant nuclear features relevant to processes from fusion to fission.
Historically, in cross section calculations of a nuclear reaction, the effects of the intrinsic structure of the reaction fragments (or clusters)  have been discussed through spectroscopic factors (SF),  determined as normalization coefficients in the reaction cross sections. However, such normalization coefficients, in general, reflect any difference between the few-body theory and experiment. Alternatively, a more direct probe of the structure can be proffered by microscopic SF that are calculated from the underlying many-body theory (e.g., as done in Ref.~\cite{SargsyanLSMDD23}). Yet, SF, as integrals over the wavefunction, do not provide insights into the formation of alpha clusters on the surface and the role of the interior nuclear medium. Thereby, to probe $\alpha$-cluster features that emerge within nuclei from  the underlying nucleon-nucleon interaction derived in the framework of chiral effective filed theory (EFT) \cite{BedaqueVKolck02,EpelbaumNGKMW02,EntemM03,Epelbaum06} (also called ``chiral potential"), here we utilize a knock-out (p, p$\alpha$) reaction, which is very sensitive to the details of alpha clustering of the target nucleus~\cite{Yoshida19}.

$\alpha$-Cluster substructures in nuclei are extremely challenging to model, since the wavefunctions expand to large distances. This is especially difficult for \textit{ab initio} approaches that utilize nucleon degrees of freedom and 
%chiral 
realistic interactions between the nucleons, typically, chiral potentials.
%inter-nucleon potentials. 
Pioneering \textit{ab initio} studies that have addressed this problem include the nuclear lattice effective field theory,
%(NLEFT), 
used to study the energy of the Hoyle state and $\alpha$-$\alpha$ scattering relevant to astrophysics \cite{EpelbaumKLM11, ElhatisariLRE15, ElhatisariEKL2017}, Green's function Monte Carlo 
%(GFMC) 
technique, which has been applied to the $\alpha$-cluster structures of $^8$Be and $^{12}$C \cite{WiringaPCP2000, CarlsonGPPSS2015}, and the hyperspherical harmonics 
%(HH) 
method, which has been utilized for analyzing giant resonance modes in $^4$He \cite{BaccaBLO13} (see, for example, the review \cite{FreerHKLM18}). Furthermore, recent advancements include $\alpha$-capture reaction descriptions in the intermediate-mass region using  wavefunctions from the symmetry-adapted no-core shell model (SA-NCSM) \cite{DreyfussLESBDD20}, studies of $\alpha$ clustering in even-even  Be and $^{12}$C isotopes using the no-core Monte Carlo shell model \cite{Otsuka22}, $\alpha+\alpha$ calculations using the no-core shell model with continuum \cite{KravvarisQHN2024}, as well as studies of the role of $\alpha$ channels using explicit continuum degrees of freedom
\cite{mercennemp19,FernandezMPM2023}.

In this letter, we provide the first \textit{ab initio} informed $\alpha$ knock-out reaction in the intermediate-mass region. In particular, we use the wavefunction of the target nucleus $^{20}$Ne, calculated in the \textit{ab initio} SA-NCSM, reviewed in Refs.~\cite{LauneyDD16,LauneyMD_ARNPS21} and detailed below.
%which 
This wavefunction accounts for the full microscopic structure of the target and when coupled to the continuum provides an $\alpha+^{16}$O cluster wavefunction. We show that this,  
%in combination with 
%proton and alpha optical potentials that enter 
when used in the theory of the reaction dynamics described below,
%few-body theory,
 yield a triple differential cross section for $^{20}$Ne(p, p$\alpha$)$^{16}$O  that is in a remarkable agreement with the data. By doing so, we gain insights into the interior and surface $\alpha$ clustering in $^{20}$Ne from first principles that are probed at different proton kinetic energies.

At incident energies of about 100 to 200 MeV, the distorted wave impulse approximation (DWIA) has been shown to be well suited for the analysis of (p, p$\alpha$) reactions \cite{Chant77}. These DWIA calculations, in which the incoming and outgoing plane waves are distorted as a result of the interaction between the reaction fragments, rely on the availability of an $\alpha$ cluster wavefunction.
%spectroscopic amplitudes (also called ``reduced width amplitudes''). 
Historically, the $\alpha$ cluster wavefunctions
%spectroscopic amplitudes 
%necessary in these DWIA calculations 
have been obtained from simple Woods-Saxon potentials that are adjusted to reproduce $\alpha$ separation energies.  In more recent studies \cite{Yoshida19}, the $^{20}$Ne(p, p$\alpha$)$^{16}$O cross sections have been calculated  with the input $\alpha$ cluster wavefunctions
%spectroscopic amplitudes 
obtained from the antisymmetrized molecular dynamics (AMD) method \cite{KanadaH1995,KanadaEnyoHO1995,KanadaEnyo98}. Here, we utilize a parameter-free input of 
%spectroscopic amplitudes 
$\alpha$ cluster wavefunctions derived from  SA-NCSM large-scale calculations for $^{20}$Ne that use a chiral potential.
%ppa reactions have been commonly studied using DWIA. describe why. In earlier works this reaction was studies with SA input from WS bound states and more recently SAs calculated in AMD were used.
\\

%\red{Discuss successful cluster approaches, NLEFT; capability of SA-NCSM}

\section{Theoretical Framework}

\subsection{\textit{Ab initio} SA-NCSM with continuum}
We employ the \textit{ab initio} symmetry-adapted no-core shell model (SA-NCSM) \cite{LauneyDD16,DytrychLDRWRBB20,LauneyMD_ARNPS21}, which treats all nucleons in the system at an equal footing and
%that 
admits chiral EFT interactions between the nucleons. The use of chiral EFT interactions, which are consistent with the chiral symmetry- and symmetry-breaking patterns of quantum chromodynamics,  enables nuclear calculations informed by physics of two  and three nucleons only.  In addition, the symmetry-adapted (SA) basis in the many-body SA-NCSM approach provides solutions to otherwise infeasible ultra-large model spaces~\cite{LauneyDD16, DytrychLDRWRBB20}, which are imperative for the description of challenging features in nuclei, such as clustering, collectivity, and coupling to the continuum.  
We note that the SA-NCSM results exactly reproduce those of the traditional NCSM \cite{NavratilVB00,Barrettnv13} for the same nuclear interaction. However, 
%by utilizing 
it utilizes
the emergent symplectic symmetry \SpR{3}$\supset$\SU{3}$\supset$\SO{3} in nuclei \cite{DytrychLDRWRBB20}, where each \SpR{3}-preserving subset of basis states describes a nuclear shape and each \SU{3}-preserving subset informs the specific deformation. By doing so, the SA-NCSM can achieve convergence of results and describe localized $\alpha$ clusters with only a fraction of the corresponding NCSM space~\cite{DreyfussLESBDD20,PhysRevLett.128.202503}.
%the SA-NCSM can use only a fraction of the corresponding complete NCSM space to achieve convergence of results and describe localized $\alpha$ clusters

Similarly to the traditional NCSM, 
the SA-NCSM uses a harmonic oscillator (HO) single-particle basis with frequency \hw~and a model space cutoff (given by the maximum  total HO excitation quanta above the valence-shell  configuration), which inform the resolution and the space size in which the nucleus resides.  The calculations become independent of \hw~ in the infinite-size model space (infinitely many HO shells), providing a parameter-free \textit{ab initio} prediction.

The cluster wavefunction, the so-called spectroscopic amplitude (also called ``reduced width amplitudes''), for the $^{20}$Ne ground state is calculated through its overlap 
%(called spectroscopic overlap) 
with an $\alpha+^{16}$O cluster state $\ket{\Psi_{ \mathfrak{a}'\mathfrak{a}''(I'^{\pi'}I''^{\pi''}); I n_r \ell}}$, which is fully antisymmetrized and normalized, as detailed in Refs. \cite{DreyfussLESBDD20,SuzukiH86}:
\begin{align}
    \label{eq:SAwithOverlap}
    u_{\nu I \ell}^{J^{\pi}}(r)
        =
        \sum_{n_r}  R_{n_r \ell}(r)\braket{\Psi_{\mathfrak{a}}^{ J^\pi (M)}}{\Psi_{ \mathfrak{a}'\mathfrak{a}''(I'^{\pi'}I''^{\pi''}); I n_r \ell} ^{J^{\pi}(M)}},
\end{align}
given for distance $r$ between the center-of-mass (c.m.) of the clusters. In Eq.~\eqref{eq:SAwithOverlap}, the cluster system is defined for a channel $\nu$, which is given by the spin and parity of each of the clusters $\nu  = \{ \mathfrak{a}; \mathfrak{a}', I'^{\pi'}, \mathfrak{a}'', I''^{\pi''}\}$ (the labels $\mathfrak{a}$, $\mathfrak{a}'$ and $\mathfrak{a}''$ denote all other quantum numbers needed to fully characterize their respective states), and a partial wave $\ell$, or the orbital angular momentum of the relative motion of the clusters, and has a good total angular momentum and parity, $J^{\pi}$, given by the coupling of the total angular momentum $I$ of the clusters to $\ell$. 

$R_{n_r \ell}(r)$ are the HO radial functions, with $n_r$ denoting the radial HO quantum number (we note that for overlaps calculated in the SA-NCSM, $R_{n_r \ell}$ are defined as positive at infinite distance).
In this study, $\nu=\{^{20}{\rm Ne}(0^+_{\rm gs});\alpha(0^+_{\rm gs}),I'^{\pi'}=0^+, ^{16}{\rm O}(0^+_{\rm gs}),I''^{\pi''}=0^+\}$, $I=0$, and $J^\pi=0^+$ [to simplify notations, $u_{\ell} (r)$ will be used henceforth].

Within a no-core shell-model framework, the 
%$A$-particle state 
$\ket{\Psi_{^{20}{\rm Ne}(0^+_{\rm gs})}^{ J^\pi (M)}}$ state of the $^{20}$Ne composite system  is calculated for particle laboratory coordinates. We note that, as detailed in Ref. \cite{DreyfussLESBDD20}, this state
is exactly factorized to intrinsic and c.m. wavefunctions. The c.m. %contribution 
wavefunction
is in the lowest HO energy and entirely integrated out in \eqref{eq:SAwithOverlap}, yielding a translationally invariant cluster wavefunction. 
%In addition, the cluster state in \eqref{eq:SAwithOverlap} is  normalized, that is, there is full antisymmetrization among the clusters.

At large  distance $r$, the cluster wavefunction  $u_{\ell}(r)$ for bound states should asymptotically approach the exact Coulomb Whittaker function, $W_{-\eta, \ell+\frac{1}{2}}$:
\begin{equation}
  r u_{\ell}(r) \rightarrow C_{ \ell}  W_{-\eta, \ell+\frac{1}{2}}(2kr),
  \label{eq:ANC}
\end{equation}
where $k=\sqrt{2\mu B}/\hbar$, with $B$ being the cluster separation energy and $\mu$ the reduced mass, $\eta$ is the Sommerfeld parameter, $\eta=\mu Z' Z''/\hbar^2k$, %\kdln{ cite Nunes book?} 
with $Z'$ and $Z''$ being the proton numbers of each of the clusters. The $C_{\ell}$ coefficient is called the asymptotic normalization coefficient (ANC). In this study, we use the SA-NCSM cluster wavefunction of Eq. \eqref{eq:SAwithOverlap} in the interior and the exact Coulomb function at large distances, both of which are matched using logarithmic derivatives within an R-matrix framework~\cite{DescouvemontB10}, as often done in many-body approaches (see, e.g.,\cite{BridaPW2011,KravvarisQHN2024,mercennemp19,DreyfussLESBDD20}). This, in turn, determines the ANC. While various methods exist for calculating ANCs (see, e.g., \cite{Brune_PRC66_2002,NollettW2011,Timofeyuk2010}), we use the  prescription of Ref. \cite{SargsyanLSMDD23}, which utilizes \eqref{eq:ANC}, while retaining the microscopically calculated SF, the norm of $u_{\ell}(r)$ of Eq. \eqref{eq:SAwithOverlap}. This prescription benefits from the use of SA-NCSM to calculate the interior wavefunction, as  sufficiently large model spaces are critical to achieve accurate descriptions in the interior.

In this study, the %$A$-particle
fully microscopic
wavefunction of the $^{20}$Ne ground state is calculated using the NNLO$_{\rm opt}$  NN chiral potential \cite{Ekstrom13} in the \textit{ab initio} SA-NCSM many-body approach, which has yielded energy spectra and observables (radii, quadrupole moments, $E2$ transitions, and charge form factors) in close agreement with experiment \cite{LauneyDD16, DytrychLDRWRBB20,LauneyMD_ARNPS21}.
The NNLO$_{\rm opt}$ is fitted only to NN scattering phase shifts and properties of the deuteron and is used without three-nucleon forces, which have been shown to contribute minimally to the three- and four-nucleon binding energies \cite{Ekstrom13}. Furthermore, the NNLO$_{\rm opt}$ NN potential has been found to reproduce various observables and yield results equivalent to those obtained from chiral potentials that require three-nucleon forces. These observables include the $^4$He electric dipole polarizability \cite{BakerLBND20}; the challenging analyzing power for elastic proton scattering on $^4$He, $^{12}$C, and $^{16}$O \cite{BurrowsEWLMNP19}; along with  B(E2) transition strengths  for $^{21,28}$Mg and $^{21}$F  \cite{Ruotsalainen19,PhysRevC.100.014322} in the SA-NCSM without 
effective charges, as well as collective and dynamical observables for $^{40}$Ca \cite{burrows2024,physrevc.110.034605} and $^{48}$Ti \cite{LauneyMD_ARNPS21}.

\subsection{Knock-out reaction theory}
\label{DWIA}
We follow the distorted-wave impulse approximation (DWIA)~\cite{Chant77,Chant83,Wakasa17} reaction analysis for $^{20}$Ne(p,p $\alpha)^{16}$O of Ref.~\cite{Yoshida19}.
Namely, the triple-differential cross section
 %The TDX 
 with respect to the kinetic energy of the emitted proton $T_p^\mathrm{L}$, emission angle of the proton $\Omega_p^\mathrm{L}$, and the $\alpha$ emission angle $\Omega_\alpha^\mathrm{L}$ is given by
\begin{equation}
 \frac{d^3\sigma}{dT_p^\mathrm{L}d\Omega_p^\mathrm{L}d\Omega_\alpha^\mathrm{L}} 
 = F_{\mathrm{kin}}C_0 \frac{d\sigma_{p\alpha}}{d\Omega_{p\alpha}}\left|\bar{T}\right|^2,
 \label{eq:dwia-tdx}
\end{equation}
where the superscript $\mathrm{L}$ indicates that the quantities are evaluated in the laboratory frame of the reaction. $F_{\mathrm{kin}}$ consists of the phase volume factor and the Jacobian from the center-of-mass frame to the laboratory frame. $d\sigma_{p\alpha}/d\Omega_{p\alpha}$ is the $p$-$\alpha$ elementary cross section.  A constant $C_0$ contains physics constants and factors related to the p-$\alpha$ elementary process.
The TDX~\eqref{eq:dwia-tdx} depends on the so-called reduced transition matrix $\bar T$ of the DWIA framework using the cluster wavefunction $u_0$, and similarly to Eq.~(6) 
of Ref.~\cite{Yoshida19} is given by
\begin{equation} 
 \bar{T} = \int d\bm{r}\, F(\bm{r}) u_{\ell=0}(r) Y_{00}(\hat{\bm{r}}),
 \label{eq:dwia-barT}
\end{equation}
assuming $\ell=0$ based on the single-peak shape of the experimental TDX.  $F(r)$ is determined by the incoming and outgoing distorted waves as defined in Eq. (7) in Ref.~\cite{Yoshida19}.

The DWIA calculations are performed using {\sc pikoe}~\cite{Ogata23}. To provide a level of specificity for completeness, we note that except the cluster wavefunction, all the inputs for the DWIA calculations are the same as in Ref.~\cite{Yoshida19}. In particular, the interactions between all pairs of reaction fragments, called optical potentials, are adopted as following:
For the proton scattering, the EDAD1 global optical potential 
of the Dirac phenomenology~\cite{Hama90,Cooper93,Cooper09} is adopted and the $\alpha$-$^{16}$O optical potential by Michel~\cite{Michel83} is adopted for the $\alpha$ scattering wave in the final state. 
We also show results that use the global proton optical potential by Koning and Delaroche~\cite{Koning03}.
The $p$-$\alpha$ elementary cross section, $d\sigma_{p\alpha}/d\Omega_{p\alpha}$, is calculated by the folding model~\cite{Toyokawa13} using the Melbourne $g$-matrix interaction.
See Sec. II of Refs.~\cite{Yoshida19,Wakasa17} for more details of the reaction framework and Eqs.~\eqref{eq:dwia-barT} and \eqref{eq:dwia-tdx}.
\begin{figure}[th]
    \centering
    {\includegraphics[width=\columnwidth]{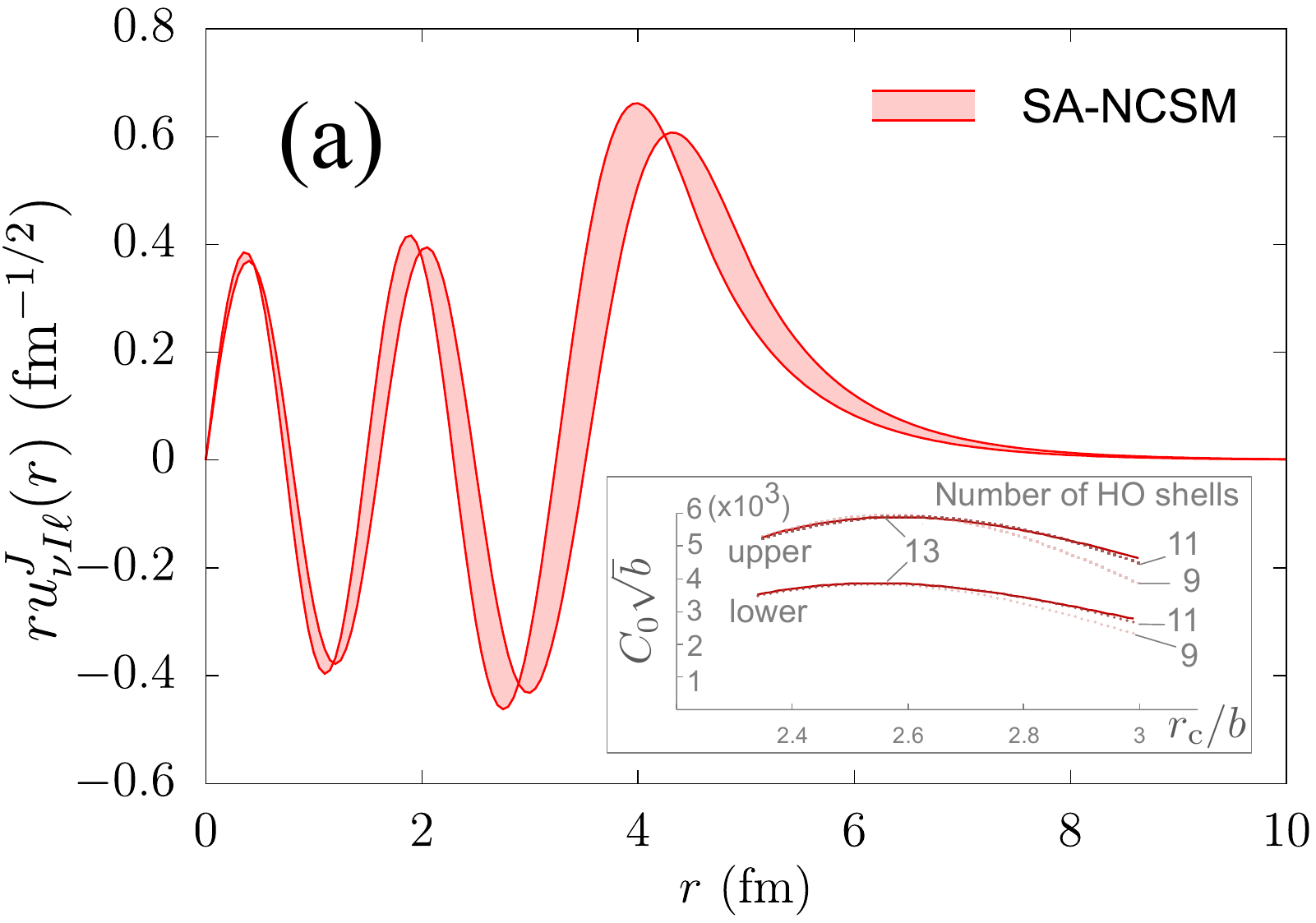}}\\
    {\includegraphics[width=\columnwidth]{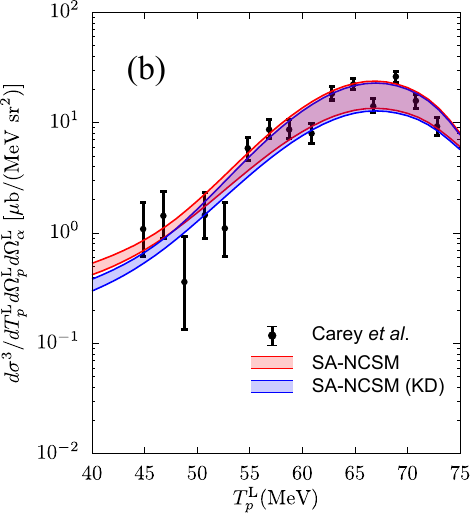}}\\
    \caption{(a) 
    %Spectroscopic amplitudes for the 
    $\alpha+^{16}${O} $S$ partial wave for the $^{20}$Ne ground state, calculated in the \textit{ab initio} symmetry-adapted no-core shell model  with the NNLO$_{\rm opt}$ NN chiral potential in 13 HO shells, with many-body model uncertainties (red band). Inset: ANC $C_0$ vs. the channel radius $r_{\rm c}$ (relative to the HO length $b$), shown for model spaces of 9, 11, and 13 HO shells.
     (b) Triple differential cross sections for the $\alpha$ knock-out reaction $^{20}$Ne(p, p$\alpha$)$^{16}$O vs. the proton kinetic energy using the cluster wavefunction shown in (a), and compared to the experiment of Ref.~\cite{Carey84} (labeled as ``Carey et al."). 
     ``SA-NCSM" shows the result using the EDAD1 optical potential for the proton scattering (red band), while ``SA-NCSM (KD)" uses the Koning-Delaroche optical potential (blue band). See text for details.
    }
    \label{fig:xsecSANCM}
\end{figure}

In Sec.~\ref{sec:result}, the $^{20}$Ne(p, p$\alpha$)$^{16}$O TDX 
%using the present spectroscopic amplitude 
is compared with the experimental data by Carey et al.~\cite{Carey84}.
In the experiment, the emission angle of the proton is fixed at $\theta_p^\mathrm{L} = 70^\circ$ and $\phi_p^\mathrm{L} = 180^\circ$, and the $\alpha$ emission angle is fixed at $\theta_\alpha^\mathrm{L} = 46.3^\circ$ and $\phi_\alpha^\mathrm{L} = 0^\circ$. The theoretical calculations are performed according to the experimental setup.
With this kinematics, the TDX distribution with respect to the kinetic energy of the emitted proton $T_p$ is measured.
At around $T_p = 67$~MeV, the so-called recoil-less condition is achieved and the TDX is peaked, reflecting the $\ell = 0$ nature of the $\alpha$ cluster.

\section{Results and Discussions}
\label{sec:result}

We report on the first \textit{ab initio} informed $\alpha$ knock-out reaction for the intermediate-mass $^{20}$Ne nucleus, which yields a triple differential cross section for 
$^{20}$Ne(p, p$\alpha$)$^{16}$O that reproduces the experimental data of Ref.~\cite{Carey84} (Fig.~\ref{fig:xsecSANCM}). The NNLO$_{\rm opt}$ chiral potential is used in the SA-NCSM many-body approach to calculate 
the $^{20}$Ne $0^+$ ground state, which in turn is projected on the $\alpha+^{16}$O $S$ partial wave to compute $u_{0}(r)$
%$u_{\alpha(0^+_{\rm gs})+^{16}{\rm O}(0^+_{\rm gs}), I=0\, \ell=0}^{J^\pi=0^+}(r)$ 
according to Eq.~\eqref{eq:SAwithOverlap} (Fig.~\ref{fig:xsecSANCM}a).
%[to simplify notations, $u_{\ell=0} (r)$ will be used henceforth].  
With $u_{0}(r)$ supplied to the few-body theoretical framework of the knock-out reaction dynamics outlined in Sec. \ref{DWIA}, the agreement to data is remarkable given that the $\alpha$-cluster features in $^{20}$Ne emerge from the underlying chiral potential that is fitted to only two-nucleon systems, with no additional parameters to tweak. Furthermore, the use of the global proton optical potential by Koning and Delaroche~\cite{Koning03} yields only marginal differences at low proton kinetic energies $T_p^\mathrm{L}$, as shown in Fig.~\ref{fig:xsecSANCM}b. This makes the output being practically independent from the specific model for the proton optical potential. Based on this, the results reported henceforth utilize the EDAD1 global optical potential used in Ref.~\cite{Yoshida19}.

In particular, we adopt the $u_{0}(r)$ $\alpha+^{16}$O cluster wavefunction for the $^{20}$Ne ground state that have been reported  and detailed in Ref.~\cite{DreyfussLESBDD20}. We note that excitations of $\alpha$ are not considered given the 20.21-MeV first excited state in $^4$He. In addition, since the NNLO$_{\rm opt}$ yields a $^4$He wavefunction that is 90-95\% in the spin-zero $s$-shell configuration, using this configuration only results in an approximation that practically does not affect the final results. To provide parameter-free input to the (p, p$\alpha$) cross section, we ensure convergence of the ANC coefficients for $u_{0}(r)$ in the limit of the infinite-size model space. This is based on an important feature we find here for the ANC, namely, while the ANC coefficients slightly increase with the model space size for $\hw=11$ MeV, they decrease for $\hw=13$ MeV, providing a stringent lower and upper limit for this quantity, $C_0=3.7(7)\times 10^3$ fm$^{-1/2}$. This is consistent with the value reported in Ref. \cite{DreyfussLESBDD20} for the range of $\hw=13$-$17$ MeV in 13 HO shells, while yielding uncertainty that is smaller by a factor of two.
In addition, Ref.~\cite{DreyfussLESBDD20} has shown that with decreasing $\hw$, at about $\hw=13$ MeV  the ANCs become independent from the $r_{\rm c}$ channel radius, the radius at which Eq.~\eqref{eq:ANC} is calculated (see Fig. 5 of Ref. \cite{DreyfussLESBDD20}). Indeed, the independence from $r_{\rm c}$ is evident by the flat regions in $C_0$ across $r_{\rm c}$, shown in the inset of Fig.~\ref{fig:xsecSANCM}, which improves with larger model spaces (naturally, for $r_{\rm c}$ smaller than the effective range of the interaction $C_0$ decreases, as manifested around $r_{\rm c} \lesssim 2.5 b$, where $b = \sqrt{\hbar\over {m\Omega}}$ is the HO length, with $m$ being the average nucleon mass).

\begin{figure}[t]
    \centering
    %(a) $\alpha$+$^{16}$O $S$ partial wave\\
   % {\includegraphics[width=\columnwidth]{amp_cmp.pdf}}\\
    %(b) $^{20}$Ne(p,p $\alpha$) $^{16}$O
    %{\includegraphics[width=\columnwidth]{20Ne_ppa_SANCSM_AMD.png}}
    {\includegraphics[width=\columnwidth]{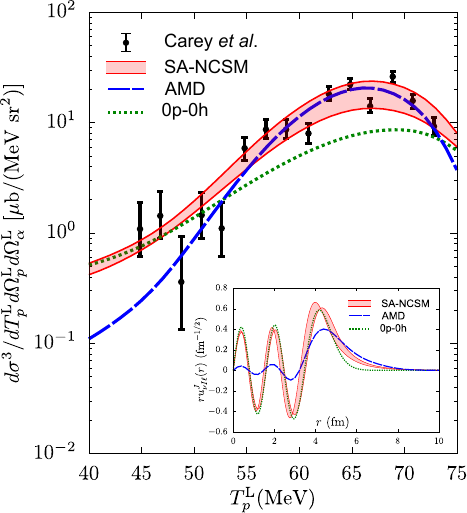}}
    \caption{Same as Fig.~\ref{fig:xsecSANCM}b, but including the cross section using the  AMD cluster wavefunction (dashed blue) and the 0p-0h contribution (dotted green) of the upper-limit SA-NCSM cluster wavefunction. Inset: same as Fig.~\ref{fig:xsecSANCM}a, but including the corresponding cluster wavefunctions.
    }
    \label{fig:xsec_cmp}
\end{figure}

Based on this estimate for the ANC, we use cluster wavefunctions calculated in the SA-NCSM in 13 HO shells for $\hw=11$ MeV as an upper limit and for $\hw=13$ MeV as a lower limit (the use of ``upper limit" reflects the ANC upper limit, but also corresponds to the upper bound in the TDX uncertainty band of Fig.~\ref{fig:xsecSANCM}b, similarly,  ``lower limit" corresponds to the ANC and TDX lower bounds). These two limits define the many-body model uncertainties. We also note that using these limits, the SA-NCSM calculations yield a $^{20}$Ne ground-state charge radius\footnote{We calculate the charge radius from the point-proton rms radius using the constants $R_p=0.84184(67)$ fm \cite{Pohl2010proton} and $R_n^2=-0.1149(27)$ fm$^2$ \cite{AngeliM2013}, plus relativistic and spin-orbit corrections as detailed in \cite{BaccaBS2012}.} 
of $2.94(9)$ fm, in agreement with the experimental value of 3.0055(21) fm \cite{AngeliM2013}. As clearly evident in Fig.~\ref{fig:xsecSANCM}a, the cluster wavefunction reveals $\alpha$ clustering around the $^{16}$O surface but also, to a lesser but comparable extent, within the $^{16}$O interior. Since wavefunctions are not observables and the $^{20}$Ne ground state ANC have not yet been deduced experimentally, the (p, p$\alpha$) data is an ideal opportunity to examine emergent clustering features.  These are also compared in Fig.~\ref{fig:xsec_cmp} to the outcomes of the successful AMD approach  and to the case of eliminating collective and continuum correlations, that is, valence-shell calculations, as discussed below.

%\kdl{The next two paragraphs are now rewritten, please check (the original text is moved to the appendix.}
The peripheral behavior of the (p, p$\alpha$) reaction, that is, the extent to which the reaction is sensitive to the surface region only, is mainly determined by the $\alpha$ emission energy, $T_\alpha^\mathrm{L}$, and the absorption effect of the emitted $\alpha$ (which is described by the imaginary part of the $\alpha$ optical potential). Since for a given 
%proton 
kinetic energy $T_p^\mathrm{L}$ of the emitted proton,  $T_p^\mathrm{L} + T_\alpha^\mathrm{L}$ are constrained by the beam energy, in addition to the $\alpha$ separation energy and the recoil of the reaction residues, higher proton kinetic energies $T_p^\mathrm{L}$ imply smaller $T_\alpha^\mathrm{L}$. Of a particular interest is the maximum of the knock-out cross section, which occurs at $T_p^\mathrm{L} \sim 67$~MeV. It corresponds to a zero recoil momentum $\bm{k}$ of the reaction residues and probes surface $\alpha$ clustering. One can understand this in the simple framework of a plane-wave (PW) approximation, where Eq.~\eqref{eq:dwia-barT} becomes $\bar{T}_{\mathrm{PW}} = \int\, d\bm{r} e^{i\bm{k}\cdot\bm{r}} u_0(\bm{r}) $ with $\left|\bar{T}_{PW}(k=0)\right|^2=\left| \int d\bm{r} \, u_0(\bm{r})   \right|^2 $, which is given by the area under the 
%\kdln{Please replace SAs by cluster wavefunctions}
cluster wavefunctions in the inset of Fig.~\ref{fig:xsec_cmp}. Since the contribution from the interior peaks practically cancels, the area is indeed determined from the surface peak. Therefore, the (p, p$\alpha$) cross section around $T_p^\mathrm{L} \sim 67$~MeV is determined by the surface region of $u_0(r)$. 

Interestingly, while the AMD approach yields overall surface clustering features and wavefunction tail similar to what the SA-NCSM with NNLO$_{\rm opt}$ reveals, the taller and more compact surface peak for SA-NCSM yields practically the same area beneath, as the shorter but wider peak in the AMD approach (Fig.~\ref{fig:xsec_cmp}).  In addition, the surface peak location for the AMD and SA-NCSM (upper limit) practically coincide ($4.4$ fm for AMD and $4.32$ fm for SA-NCSM). Further accounting for the absorption effect of the emitted $\alpha$, this results in a good agreement of both AMD and SA-NCSM approaches to the data around $T_p^\mathrm{L} \sim 67$~MeV that have small uncertainties. Furthermore, we consider only the 0-particle-0-hole (0p-0h) contribution of the SA-NCSM cluster wavefunction (upper limit) as input to the cross section, that is, an alpha particle in the valence shell in $^{20}$Ne. In this case, the narrower surface peak (see the green dotted curve in the inset of Fig.~\ref{fig:xsec_cmp}) largely underestimates the cross section. This emphasizes the critical role of correlations beyond the 0p-0h contribution such as collective correlations and coupling to the continuum that enter the entire SA-NCSM cluster wavefunction, for reproducing the cross section. 

Similarly, low $T_p^\mathrm{L}$ proton kinetic energies  probe -- to a certain extent -- higher-energy alpha particles that leave the nuclear interior. Even though the scattering waves have a smaller amplitude in the interior region due to the absorption effect arising from the imaginary part of the optical potential and suppress the reaction sensitivity to this region, the larger difference of the interior AMD and SA-NCSM wavefunctions lead to some noticeable deviations (at a logarithmic scale) at low $T_p^\mathrm{L}$ energies, in both the height and the shape of the TDX tail (Fig.~\ref{fig:xsec_cmp}). However, in this region, the 0p-0h contribution yields a cross section comparable with that of the entire SA-NCSM cluster wavefunction, likely as a result of the similar wavefunction behavior at short distances (Fig.~\ref{fig:xsec_cmp}, inset). In particular, the NNLO$_{\rm opt}$ chiral potential suggests strong mixing between the clusters within the nuclear medium predominantly arising from  0p-0h configurations, as discussed next. With such mixing, the \textit{ab initio} informed cross section agrees with the measurements below $T_p^\mathrm{L} \sim 55$ MeV within about $1\sigma$ (Fig.~\ref{fig:xsecSANCM}b).
\begin{figure}
    \centering 
    \includegraphics[width=0.49\linewidth]{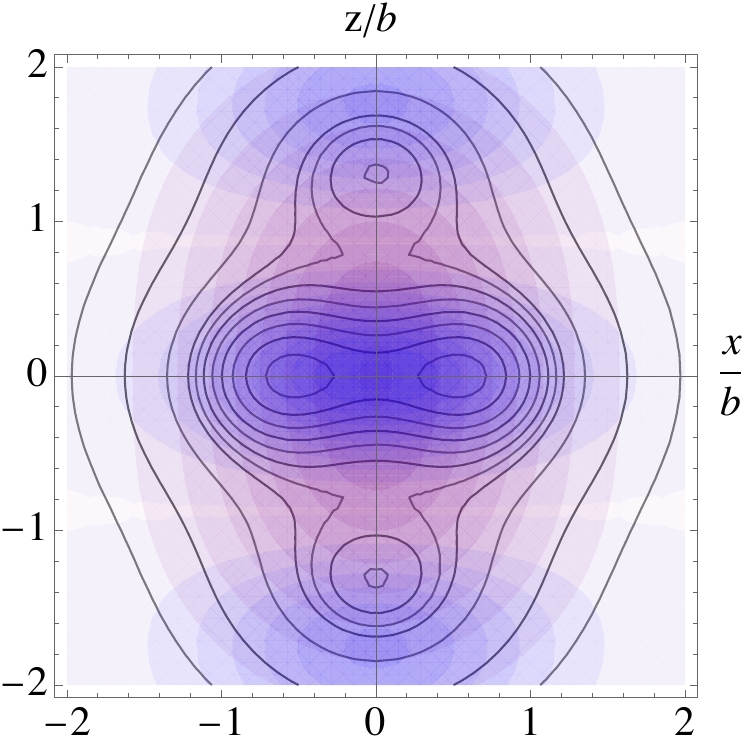}
    \includegraphics[width=0.49\linewidth]{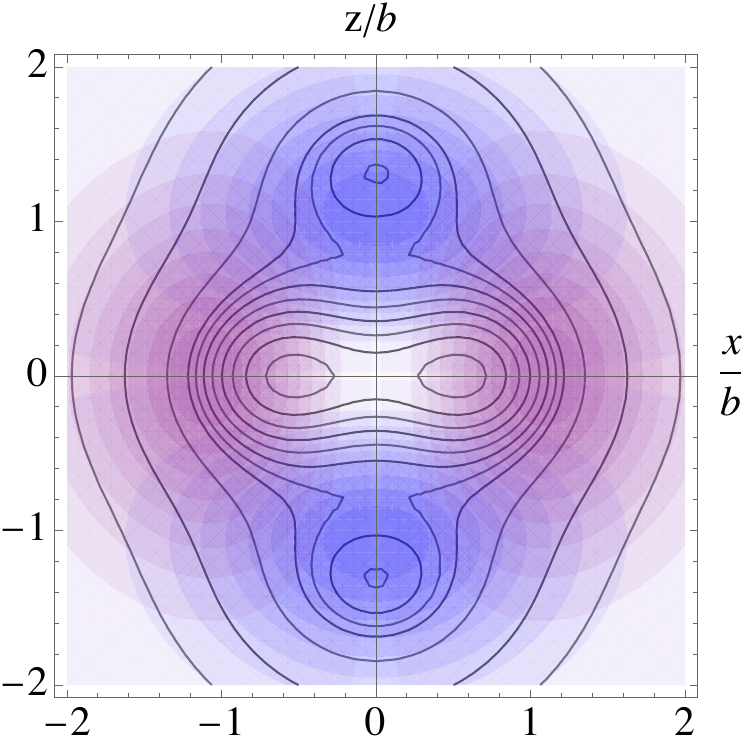} \\ (a) \hspace{1.46in} (b)
    \caption{Intrinsic one-body density profile for $^{20}$Ne (black contours) in the $\{x,z\}$ plane (relative to 
    %the HO length 
    $b$), shown with the contributions from the lowest five most dominant natural single-particle orbitals with (a) positive parity: $s$-shell dominated  (purple) and $sd$-shell dominated (blue), as well as (b) negative parity: $p$-shell dominated 
    with HO excitations mainly along the $z$ axis (blue) and $x$ axis (purple).
    }
    \label{fig:obd}
\end{figure}

To understand the cluster features within the nuclear medium, as emerging from the NNLO$_{\rm opt}$ chiral potential, we examine the intrinsic (\textit{body-fixed}) one-body density (Fig.~\ref{fig:obd}), {which informs about the probability of finding a single particle at a given position within the $^{20}$Ne ground-state wavefunction. The eigenvectors of the one-body density matrix are often referred to as natural orbitals. We find that 
%which is made of the densities of practically 
only five spatial natural orbitals are largely occupied (with maximum of four particles per orbital, two protons and two neutrons with spin up and down). Each of these natural orbitals contributes $19.9\%$ to the overall occupancy and is calculated as an eigenvector of the one-body density matrix for the SA-NCSM $^{20}$Ne ground state using basis configurations with probability amplitudes more than 2\%. 
Clearly, the natural orbitals with dominant $p$-shell and $sd$-shell contributions to the density (blue shades in Fig.~\ref{fig:obd}) are important in shaping the $\alpha$ surface peak. However, among the two, the positive-parity natural orbital contributes significantly to both the surface peak and the interior, and thus highly overlaps with the cluster in the center described by a dominant $s$-shell configuration (Fig.~\ref{fig:obd}a).

%\section{Conclusions}
In short, we present the first study that probes details of emergent clustering features from chiral potentials in the intermediate-mass region, including surface $\alpha$ clustering and cluster mixing. We show that the $\alpha+^{16}$O cluster wavefunction from the
SA-NCSM with NNLO$_{\rm opt}$, coupled with proton and alpha optical potentials that enter the few-body theory, reproduce within the statistics the triple differential cross section for the $\alpha$ knock-out reaction on $^{20}$Ne. In particular, we show that, in the region around the maximum cross section that probes surface $\alpha$ clustering, the SA-NCSM approach yields similar cluster features to those of AMD, confirming the critical need for collective correlations and coupling to the continuum for reproducing the cross section. Whereas within the nuclear medium, the NNLO$_{\rm opt}$ chiral potential suggests strong mixing between the clusters and largely overlapping clusters. The remarkable agreement to experiment further validates the properties of the NNLO$_{\rm opt}$ chiral
potential for predicting collective and clustering features in the intermediate-mass region, and confirms the significance of the \textit{ab initio} SA-NCSM approach that is capable to facilitate the relevant many-body configurations and, thereby, to describe localized clustering from first principles beyond the lightest nuclear systems. This work paves the way toward using cluster wavefunctions derived from \textit{ab initio} SA-NCSM calculations as input to $\alpha$ knock-out cross sections, as well as to inform and support   (p,p$\alpha$) measurements in the intermediate- and medium-mass region.

\section{Acknowledgments}

%\blue{(TBD; is this part of the 6-page limit?)} 
This work was supported by the U.S. Department of Energy (DE-SC0023532, DE-FG02-93ER40756), as well as in part by the U.S. National Science Foundation (PHY-2209060) and the Czech Science Foundation (22-14497S). This work is also supported in part by Grant-in-Aid for Scientific Research (No. JP20K14475, and No. JP21H04975) from Japan Society for the Promotion of Science (JSPS). 
 This material is based upon work supported by the U.S. Department of Energy, Office of Science, Office of Nuclear Physics, under the FRIB Theory Alliance award DE-SC0013617. This work was performed in part under the auspices of the U.S. Department of Energy by Lawrence Livermore National Laboratory under Contract DE-AC52-07NA27344, with support from the Laboratory Directed Research and Development Program, Projects 19-ERD-017 and 24-ERD-023. This work benefited from high performance computational resources provided by LSU (www.hpc.lsu.edu), the National Energy Research Scientific Computing Center (NERSC), a U.S. Department of Energy Office of Science User Facility at Lawrence Berkeley National Laboratory operated under Contract No. DE-AC02-05CH11231, as well as the Frontera computing project at the Texas Advanced Computing Center, made possible by National Science Foundation award OAC-1818253.

%\bibliography{lsu_latest,Ne20ppa}
\bibliography{20Ne_ppa}

\end{document}